\documentclass[12pt]{iopart}

\pdfoutput=1
\usepackage{lineno,hyperref}
\usepackage{color}
\usepackage{siunitx}
\usepackage{iopams}
\usepackage{graphicx}
\modulolinenumbers[5]

\DeclareSIUnit{\solarradius}{\ensuremath{R_\odot}}

\newcommand{\Bilgi}{Istanbul Bilgi University, High Energy Physics Research Center, Eyup, Istanbul, 34060, Turkey}
\newcommand{\Saclay}{IRFU, CEA, Universit\'{e} Paris-Saclay, Gif-sur-Yvette, France}
\newcommand{\CERN}{European Organization for Nuclear Research (CERN), Gen\`eve, Switzerland}
\newcommand{\INR}{Institute for Nuclear Research (INR), Russian Academy of Sciences, Moscow, Russia}
\newcommand{\MPE}{Max-Planck-Institut f\"{u}r Extraterrestrische Physik, Garching, Germany}
\newcommand{\infn}{Istituto Nazionale di Fisica Nucleare (INFN), Sezione di Trieste, Trieste, Italy}
\newcommand{\Trieste}{Universit\`a di Trieste, Trieste, Italy}
\newcommand{\Zaragoza}{Grupo de Investigaci\'{o}n de F\'{\i}sica Nuclear y Astropart\'{\i}culas, Universidad de Zaragoza,Zaragoza, Spain }
\newcommand{\Columbia}{Physics Department and Columbia Astrophysics Laboratory, Columbia University, New York, NY 10027, USA}
\newcommand{\DTU}{DTU Space, National Space Institute, Technical University of Denmark, Lyngby, Denmark}

\newcommand{\Thessaloniki}{Aristotle University of Thessaloniki, Thessaloniki, Greece}
\newcommand{\Demokritos}{National Center for Scientific Research ``Demokritos'', Athens, Greece}
\newcommand{\Freiburg}{Albert-Ludwigs-Universit\"{a}t Freiburg, Freiburg, Germany}
\newcommand{\Patras}{Physics Department, University of Patras, Patras, Greece}
\newcommand{\Athens}{National Technical University of Athens, Athens, Greece}

\newcommand{\Vancouver}{Department of Physics and Astronomy, University of British Columbia, Vancouver, Canada}
\newcommand{\Darmstadt}{Technische Universit\"{a}t Darmstadt, IKP, Darmstadt, Germany}
\newcommand{\Frankfurt}{Johann Wolfgang Goethe-Universit\"at, Institut f\"ur Angewandte Physik, Frankfurt am Main, Germany}
\newcommand{\Zagreb}{Rudjer Bo\v{s}kovi\'{c} Institute, Zagreb, Croatia}
\newcommand{\MPP}{Max-Planck-Institut f\"{u}r Physik (Werner-Heisenberg-Institut), M\"unchen, Germany}
\newcommand{\LLNL}{Lawrence Livermore National Laboratory, Livermore, CA 94550, USA}
\newcommand{\MPIS}{Max-Planck-Institut f\"{u}r Sonnensystemforschung, G\"{o}ttingen, Germany}

\newcommand{\Korea}{School of Space Research, Kyung Hee University, Yongin, Republic of Korea.}

\newcommand{\Rijeka}{Department of Physics and Centre for Micro and Nano Sciences and Technologies, University of Rijeka, Rijeka, Croatia.}

\newcommand{\California}{Dept.\ of Physics and Astronomy, University of California, Irvine, CA 92697, USA.}
\newcommand{\Bonn}{Physikalisches Institut, University of Bonn, Germany.}
\newcommand{\CAPP}{Center for Axion and Precision Physics Research, Institute for Basic Science (IBS), Daejeon 34141, Republic of Korea.}
\newcommand{\KAIST}{Department of Physics, Korea Advanced Institute of Science and Technology (KAIST), Daejeon 34141, Republic of Korea}
\newcommand{\ELI}{Extreme Light Infrastructure - Nuclear Physics (ELI-NP), 077125 Magurele, Romania}
\newcommand{\IHEP}{Institute of High Energy Physics, Chinese Academy of Sciences, Beijing}
\newcommand{\SaclayT}{Institut de Physique Théorique, CEA, IPhT, CNRS, URA 2306, F-91191Gif/Yvette Cedex, France}
\newcommand{\jiaotong}{Xi'An Jiaotong University, School of Science, Xi'An, P.R. China}
\newcommand{\yale}{Department of Physics, Yale University, New Haven, Connecticut 06511, USA}

\begin{document}

\title{Improved Search for Solar Chameleons with a GridPix Detector at CAST}

\author{V.~Anastassopoulos$^{1}$, S.~Aune$^{2}$, K.~Barth$^{3}$, A.~Belov$^{4}$, H.~Br\"auninger$^{5}$, G.~Cantatore$^{6,7}$, J.~M.~Carmona$^{8}$, J.~F.~Castel$^{8}$, S.~A.~Cetin$^{9}$, F.~Christensen$^{10}$, T.~Dafni$^{8}$, M.~Davenport$^{3}$, A.~Dermenev$^{4}$, K.~Desch$^{11}$, B.~D\"obrich$^{3}$, C.~Eleftheriadis$^{12}$, G.~Fanourakis$^{13}$, E.~Ferrer-Ribas$^{2}$, H.~Fischer$^{14}$, W.~Funk$^{3}$, J.~A.~Garc\' ia$^{8}$\footnote{Present address: \IHEP}, A.~Gardikiotis$^{1}$, J.~G.~Garza$^{8}$, E.~N.~Gazis$^{15}$, T.~Geralis$^{13}$, I.~Giomataris$^{2}$, S.~Gninenko$^{4}$, C.~J.~Hailey$^{16}$, M.~D.~Hasinoff$^{17}$, D.~H.~H.~Hoffmann$^{18}$, F.~J.~Iguaz$^{8}$, I.~G.~Irastorza$^{8}$, A.~Jakobsen$^{10}$, J.~Jacoby$^{19}$, K.~Jakov\v ci\' c$^{20}$, J.~Kaminski$^{11}$, M.~Karuza$^{6\text{,}21}$, S.~Kostoglou$^{3}$, N.~Kralj$^{21}$\footnote{Present address: \yale}, M.~Kr\v{c}mar$^{20}$, C.~Krieger$^{11}$\footnote{Present address: Institute of Experimental Physics, University of Hamburg, Germany}, B.~Laki\'{c}$^{20}$, J.~M.~Laurent$^{3}$, A.~Liolios$^{12}$, A.~Ljubi\v{c}i\'{c}$^{20}$, G.~Luz\'on$^{8}$, M.~Maroudas$^{1}$, L.~Miceli$^{22}$, S.~Neff$^{23}$, I.~Ortega$^{8\text{,}3}$, T.~Papaevangelou$^{2}$, K.~Paraschou$^{12}$, M.~J.~Pivovaroff$^{24}$, G.~Raffelt$^{25}$, M.~Rosu$^{23}$\footnote{Present address: \ELI}, J.~Ruz$^{24}$, E.~Ruiz Ch\'oliz$^{8}$, I.~Savvidis$^{12}$, S.~Schmidt$^{11}$, Y.~K.~Semertzidis$^{22\text{,}26}$, S.~K.~Solanki$^{27}$\footnote{Also at: \Korea}, L.~Stewart$^{3}$, T.~Vafeiadis$^{3}$, J.~K.~Vogel$^{24}$, M.~Vretenar$^{21}$, W.~Wuensch$^{3}$, S.~C.~Yildiz$^{9}$\footnote{Present address: \California} and K.~Zioutas$^{1\text{,}3}$ (CAST Collaboration) and P.~Brax$^{28}$}

\address{$^{1}$ \Patras}
\address{$^{2}$ \Saclay}
\address{$^{3}$ \CERN}
\address{$^{4}$ \INR}
\address{$^{5}$ \MPE}
\address{$^{6}$ \infn}
\address{$^{7}$ \Trieste}
\address{$^{8}$ \Zaragoza}
\address{$^{9}$ \Bilgi}
\address{$^{10}$ \DTU}
\address{$^{11}$ \Bonn}
\address{$^{12}$ \Thessaloniki}
\address{$^{13}$ \Demokritos}
\address{$^{14}$ \Freiburg}
\address{$^{15}$ \Athens}
\address{$^{16}$ \Columbia}
\address{$^{17}$ \Vancouver}
\address{$^{18}$ \jiaotong}
\address{$^{19}$ \Frankfurt}
\address{$^{20}$ \Zagreb}
\address{$^{21}$ \Rijeka}
\address{$^{22}$ \CAPP}
\address{$^{23}$ \Darmstadt}
\address{$^{24}$ \LLNL}
\address{$^{25}$ \MPP}
\address{$^{26}$ \KAIST}
\address{$^{27}$ \MPIS}
\address{$^{28}$ \SaclayT}

\ead{krieger@physik.uni-bonn.de}

\begin{abstract}

We report on a new search for solar chameleons with the CERN Axion Solar Telescope (CAST). A GridPix detector was used to search for soft X-ray photons in the energy range from \SI{200}{\eV} to $\SI{10}{\keV}$ from converted solar chameleons. No significant excess over the expected background has been observed
in the data taken in 2014 and 2015. We set an improved limit on the
chameleon photon coupling, $\beta_{\gamma}\lesssim\num{5.7e10}$ for
$\num{1}<\beta_\text{m}<\num{e6}$ at \SI{95}{\percent} C.L. improving our previous results by a factor two and for the first time reaching sensitivity below the solar luminosity bound for tachocline magnetic fields up to \SI{12.5}{\tesla}.
\end{abstract}

\submitto{\JCAP}
\maketitle


\section{Introduction}

Dark energy in particular, as well as the dark sector of cosmology in general, is one of today's great challenges in fundamental physics. Dark energy, needed to explain the observed acceleration of the universe's expansion, could be introduced by modifying General Relativity. This can be accomplished via a new scalar field along with a screening mechanism to avoid unnatural effects such as as the appearance of a fifth force with long range. In case of the chameleon, one of the leading candidates for dark energy, non-linear self-interaction and interactions with matter cause these particles to have an "effective mass" depending on the ambient mass (energy) density. This leads to the so-called chameleon screening mechanism, giving rise to the name of this dark energy candidate~\cite{khoury2004,khoury2004a,brax2004} (for a comprehensive theoretical treatment we refer to~\cite{joyce2015}). Experimental constraints on chameleon models arise e.g. from fifth force gravity experiments, atomic spectroscopy, and atom and neutron interferometry. A recent summary of experimental constraints is given in~\cite{burrage2018}.

Although the chameleon currently can only be described as a low energy model in the form of an effective field theory, missing a high energy description from an ultraviolet completion (e.g. string theory), its introduction predicts interesting phenomena; this justifies investigations and searches like the one presented here.

Through an effective chameleon-photon coupling chameleons can be produced via the Primakoff effect, similar to axions, leading to the prediction of solar chameleons. While for chameleon production in the nuclear coulomb fields of the solar core's plasma no theoretical calculations exist up to now, one can consider regions in the solar interior featuring strong transverse magnetic fields. The tachocline, a solar region located at approximately \SI{0.7}{\solarradius}, is widely believed to be a source of highly intense magnetic fields formed through differential rotation. The chameleon production in the tachocline and their propagation through the solar medium as well as their journey to an earth-based helioscope have been studied in detail in the proposal~\cite{brax2010} and also in~\cite{brax2012}.

Solar chameleons from the tachocline have energies below \SI{2}{\keV} (the flux typically peaking at about \SI{600}{\eV}). Through the inverse Primakoff effect they can be (re)converted into soft X-ray photons inside a strong magnetic field. Thus, an axion helioscope such as the CERN Axion Solar Telescope (CAST)~\cite{zioutas1999} can be used as a chameleon helioscope, given X-ray detectors sensitive in the sub-\SI{}{\keV} energy range.

As the effective mass of the chameleon decreases with falling density, the free space from Sun to Earth and the Earth's atmosphere are traversed basically unhindered, so are the evacuated cold bores of the CAST magnet. Inside detector materials, especially those of classical dark matter experiments, the effective chameleon mass is large. If the effective mass exceeds the chameleon's energy, it cannot enter the corresponding material but is reflected off. In case its energy is much larger than the effective mass inside a certain material, the chameleon will hardly interact with the material, making it difficult to detect in general.

In 2013 a first search for chameleons at CAST was performed using a silicon drift detector built mostly from commercially available components~\cite{anastassopoulos2015} following the proposal of the chameleon helisocope technique in~\cite{brax2010}. In this paper we present the results of a follow-up of this first search utilizing the powerful combination of an X-ray telescope~\cite{kuster2007} with a novel GridPix detector~\cite{krieger2013,krieger2017} improving on our previous result by about a factor of two and for the first time reaching sensitivities below the solar luminosity bound. Also, a different technique exploiting the chameleon matter coupling by directly detecting solar chameleons focused by an X-ray telescope on a radiation pressure device~\cite{baum2014} is pursued by CAST and its results will be published elsewhere.

Here, we discuss uncertainties in the tachocline magnetic field strength, its radial position and dimension as well as the fraction of solar luminosity emitted as chameleons. We also consider different cases of the chameleon potential and show that our upper bound, $\beta_\gamma\lesssim\num{5.7e10}$, is almost independent of the type of inverse power law potential used.

\section{Theoretical solar chameleon spectrum}

Solar chameleons can be produced through the Primakoff effect from the photon flux emanating from the solar core~\cite{brax2010,brax2012}. In the following is given a (very) brief summary of the computation and calculation of the solar chameleon flux emerging from the solar tachocline due to photon-to-chameleon conversion inside the strong tachocline magnetic field; for more details the reader is referred to~\cite{anastassopoulos2015} and especially~\cite{brax2012}.

The conversion probability of photons into chameleons within a magnetized region with constant magnetic field $B$ over a distance $l$ can be given as~\cite{brax2012}
\begin{equation}
p_{\gamma\to\phi}(l)= \frac{\beta_\gamma^2 B^2l_\omega^2}{4 M_\text{Pl}^2 }\sin^2 \frac{l}{l_\omega}\ ,
\label{prop}
\end{equation}
where $M_\text{Pl}\sim\SI{2e18}{\GeV}$ is the reduced Planck mass, the coherence length $l_\omega$ is given by
\begin{equation}
l_\omega= \frac{4\omega}{m^2_\text{eff}}
\end{equation}
with the effective chameleon mass
\begin{equation}
m^2_\text{eff}=\beta_\text{m}^{(n+2)/(n+1)} \omega_\rho^2 -\omega_\text{pl}^2\ ,
\end{equation}
where we have defined 
\begin{equation}
\omega_\rho^2= \frac{(n+1) \rho}{M_\text{Pl}} (\frac{\rho}{n\,M_\text{Pl}\Lambda^{n+4}})^{1/(n+1)}
\end{equation}
and the plasma frequency is $ \omega_\text{pl}^2 = \frac{4\pi \alpha\rho}{m_em_p}$. Here, $\alpha\!\sim\!1/137$ is the fine structure constant, $m_p$ and $m_e$ are the proton and electron mass respectively. The effective chameleon mass depends on the density $\rho$ and the chameleon matter coupling constant $\beta_\text{m}$. The index $n>0$ defines the chameleon model and is linked to the scalar potential $\frac {\Lambda^{n+4}}{\phi^n}$ where $\Lambda\sim\SI{e-3}{\eV}$ is the dark energy scale. Here, we assume the mixing angle $\theta= \frac{\omega B \beta_\gamma}{M_\text{P} m^2_\text{eff}}$ to be $\theta\lesssim\num{1}$.

Taking into account the random walk that photons perform within the solar plasma and integrating over the Sun or the tachocline region, it is possible to compute the total chameleon luminosity of the the Sun. Assuming non-resonant chameleon production, thus restricting the chameleon matter coupling to $\num{1}<\beta_\text{m}<\num{e6}$, both the solar chameleon flux as well as the total chameleon luminosity of the Sun depend on $\beta_\gamma^2$. Calibrating $\beta_\gamma$ in a way so the solar chameleon luminosity does not exceed \SI{10}{\percent} of the solar luminosity the upper solar luminosity bound $\beta_\gamma^\text{sun}$ can be computed to $\beta_\gamma^\text{sun}=\num{6.46e10}$ for $n=1$, a tachocline of width \SI{0.01}{\solarradius} located at \SI{0.7}{\solarradius} and a tachocline magnetic field of \SI{10}{\tesla} ~\cite{anastassopoulos2015}.

\section{Experimental setup}

As the inverse Primakoff effect for chameleons is quite similar to the same effect for axions, an axion helioscope such as CAST can be used as a chameleon helioscope. The axion-photon coupling constant $g_{a\gamma}$~\cite{anastassopoulos2017} is then replaced by the chameleon-photon coupling constant $\beta_\gamma$. Due to the different production region in the Sun, the X-ray spectrum expected from solar chameleons converted into photons peaks around \SI{600}{\eV} instead of a few \SI{}{\keV} as it does for axions. Thus, X-ray detectors sensitive in the sub-\SI{}{\keV} range are required as well as fully evacuated cold bores.

For the solar axion search the CAST detectors were matched in terms of sensitivity and energy threshold to the expected solar axion spectrum, meaning an energy threshold of above \SI{1}{\keV}. Additionally, after CAST's phase~I operation with evacuated cold bores in 2003 and 2004~\cite{zioutas2005,andriamonje2007}, the cold bores were filled with helium between 2005 and 2012~\cite{arik2009,arik2011,arik2014}. When CAST switched back to vacuum operation in 2013, the thin cold windows inside the bores that showed a cutoff at \SI{1}{\keV} were removed and one of the X-ray detectors of the experiment, the pnCCD~\cite{kuster2007}, was decommissioned. This allowed for the installation of a new X-ray detector with sub-\SI{}{\keV} sensitivity, converting CAST into a chameleon helioscope.

The new X-ray detector, the GridPix detector~\cite{krieger2017}, was installed in October 2014 behind the MPE X-ray telescope (XRT) of CAST~\cite{kuster2007}, a flight spare of the ABRIXAS space mission, and took data until it was dismantled at the end of 2015. While the MPE XRT was being (re)calibrated at the PANTER facility, a first chameleon search was conducted at CAST during a short period in 2013 using a silicon drift detector (SDD) built from mostly commercially available equipment~\cite{anastassopoulos2015}. This first result of CAST's chameleon search is now improved by the powerful combination of the GridPix detector with the X-ray optics.

The GridPix detector was mounted to the XRT by means of a small interface vacuum system as described in~\cite{krieger2017}. The detector and its infrastructure were located on the sunrise side of CAST, thus participating in the morning solar trackings (about \SI{90}{\minute} per day). Solar chameleons could convert into soft X-ray photons within the cold bore (diameter \SI{43}{\milli\metre}) inside the magnetic field of \SI{9}{\tesla} provided by the CAST magnet (a decommissioned LHC prototype magnet) over a length of \SI{9.26}{\metre} and then be detected in the XRT/GridPix system.

\section{The GridPix detector}
A gaseous X-ray detector is used to detect the photons focused by the
XRT. This detector and its performance during the data taking are described in
detail in~\cite{krieger2017}, while here is given only a brief
summary. The basis of the detector is a GridPix, which is a
combination of a Timepix ASIC~\cite{llopart2007} and a Micromegas gas
amplification stage. The readout chip features $\num{256}\times\num{256}$ pixels with a
pitch of $\num{55}\times\SI{55}{\micro\metre\squared}$. The Micromegas stage was
produced directly on top of the chip by photolithographic postprocessing techniques,
which allow for small feature sizes and precise alignment. In this way each
grid hole of the mesh was be placed directly above one pixel. A single
electron liberated by the X-ray photon interaction is guided into one hole,
where an avalanche multiplication increases the charge by a gain of $G\approx\num{2500}$. Since the complete avalanche is collected by one readout pixel with a
typical threshold around \num{700} electrons, signals of most primary electrons are recorded individually resulting in a high resolution image of the event. 

\begin{figure}[tbh]
\centering
\def\svgwidth{10cm}
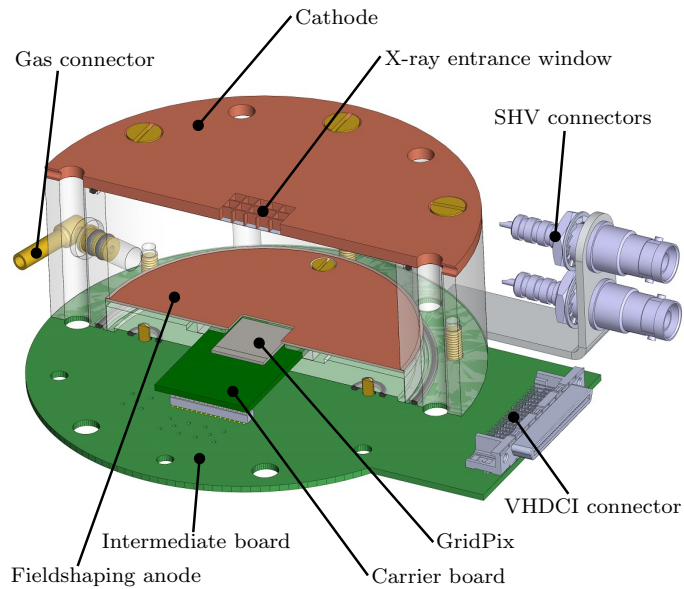
\caption{Cross-sectional view of the detector.}
\label{fig_detector}
\end{figure}

A cross-sectional view of the detector is shown in \fref{fig_detector}. The
drift volume has a length of \SI{3}{\centi\metre} and is filled with a gas mixture of
argon:isobutane (97.7:2.3) at a constant pressure of \SI{1050}{\milli\bar}. A drift field 
of \SI[per-mode=symbol]{500}{\volt\per\centi\metre} is applied.

\begin{figure}[tbh]
\centering
\includegraphics[width=.8\textwidth]{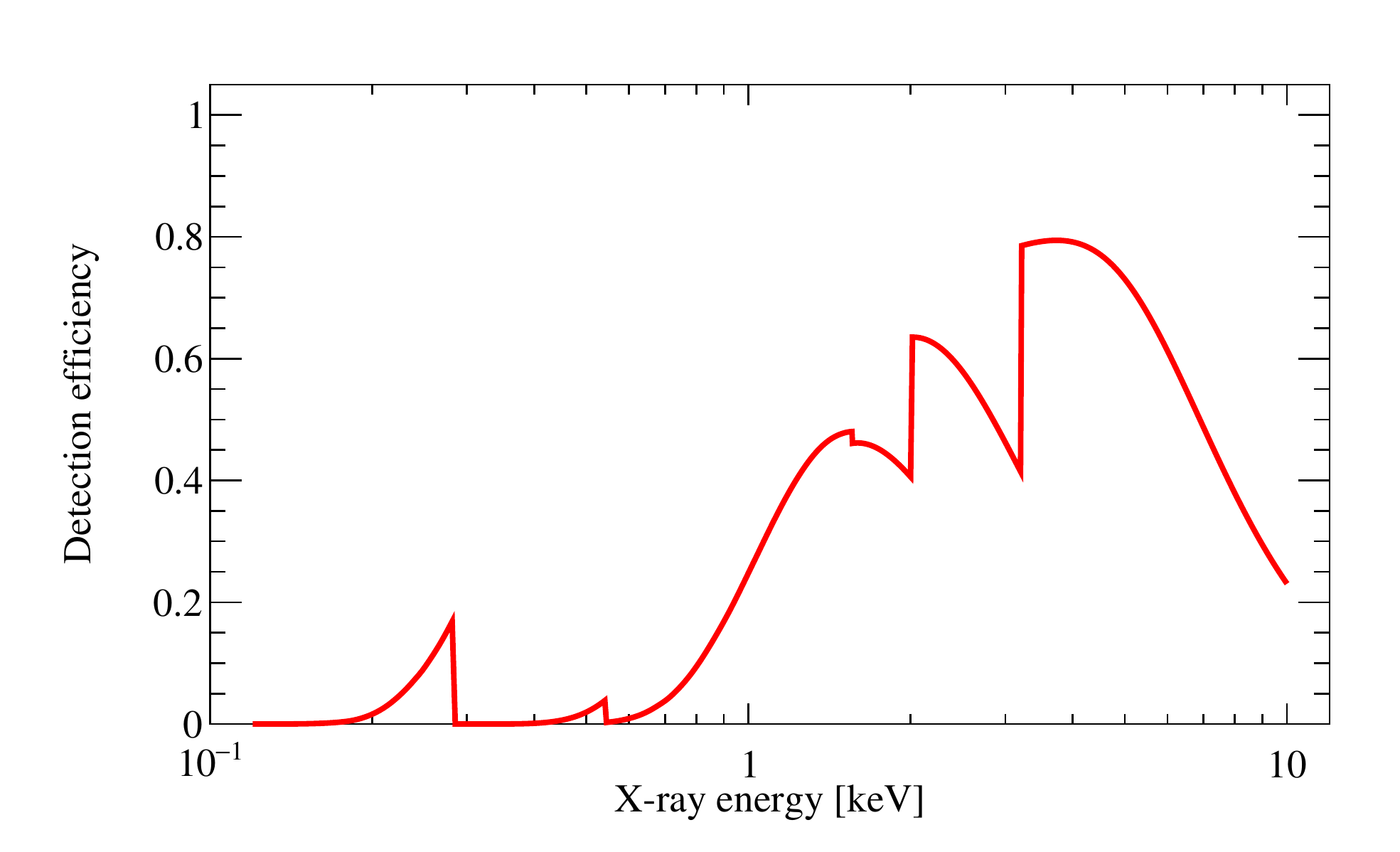}
\caption{Total detection efficiency of the detector. Curve based on transmission data provided by a web-based generator~\cite{gullikson2010} relying on the semi-empirical approach described in~\cite{henke1993}.}
\label{fig_transparency}
\end{figure}

A crucial request for the chameleon search is a high detection efficiency for
low-energy X-ray photons. Minimizing the material in front of the
sensitive detector volume is thus of high importance. Therefore, a
\SI{2}{\micro\metre} thick Mylar foil with a \SI{40}{\nano\metre} layer of
aluminum was used as entrance window and simultaneously as cathode of the
drift volume. The foil was mounted on a copper strongback covering \SI{17.4}{\percent} of
the total area. To improve the vacuum in the XRT down to the required
\SI{2e-6}{\milli\bar} a differential pumping scheme had to be used, requiring
an additional vacuum window made of \SI{0.9}{\micro\metre} of Mylar. The total
transparency of all windows multiplied by the absorption probability in the gas volume
yields the detection probability and is shown in \fref{fig_transparency} depending on the X-ray energy.

Prior to the data taking campaigns in 2014 and 2015, the detector response to X-ray photons of different energies
was characterized with a variable X-ray generator~\cite{krieger2018}. A stable operation, good
energy linearity and energy resolution were demonstrated. Also, a data
sample of signals of pure X-ray photons, important for a likelihood based background suppression were collected.

The operation of the GridPix detector at CAST lasted over six months split in a short test run in
October and November 2014 and a long run from June to November 2015. In total \SI{254}{\hour} of solar tracking and \SI{4785}{\hour} of background
data were recorded. The detector worked reliably and no
detector related downtime occurred. A daily calibration run with an
$^{55}$Fe source showed variations of only a few percent in the gas gain, which is
expected because of the changing environmental conditions, in particular of
the ambient temperature.

\section{Background suppression and rates}

All events recorded with the GridPix detector that contain at least three activated pixels are reconstructed using the MarlinTPC framework~\cite{abernathy2008}. Details on the reconstruction and data acquisition scheme can be found in~\cite{krieger2017,krieger2018}. Based on information gained from the daily in-situ calibrations with the $^{55}$Fe source, the energy of each reconstructed event is computed along with a set of eventshape variables exploiting the high spatial resolution of the Timepix ASIC. 

In~\cite{krieger2018} three eventshape variables have been identified, which provide a good separation power to differentiate X-ray events from non-X-ray events: eccentricity (a measure for the circularity of the event), length divided by root mean square (rms) along the short axis and fraction of pixels within radius of one rms (the latter two providing handles on the shape of the distribution of pixels within the reconstructed event). From the measurements at a variable X-ray generator~\cite{krieger2018} reference distributions of these variables for eight energy ranges in the regime up to \SI{10}{\keV} were obtained. These are used here to compute a likelihood value for each reconstructed event based on the reference distributions of the corresponding energy range. This provides a measure of the probability (likelihood) for an X-ray, of given energy, to look like the observed event. By only keeping those events passing a certain threshold for the likelihood value, the vast majority of non-X-ray events is removed from the data set resulting in a suppression of non-X-ray-like background events. To find the optimal likelihood threshold for each energy range the same method is applied to pure X-ray calibration datasets yielding the likelihood distribution for X-ray events of each energy range. As the likelihood distributions of X-ray and background events overlap slightly, preventing a perfect separation, the likelihood threshold is found as a compromise between high background suppression on the one hand and low loss of real X-ray events on the other. Here, the likelihood thresholds for the different energy ranges are set in a way that independently of the X-ray energy \SI{80}{\percent} of the X-ray events pass the threshold, defining a software efficiency $\epsilon_\text{SW}=\num{0.8}$. As the chosen eventshape variables are constructed to be independent of drift properties, like diffusion constant and drift distance to avoid the necessity of e.g. a temperature compensation, an additional loose cleaning cut is required to remove those X-ray-like events which due to their size (rms value) cannot stem from X-ray conversions inside the detector.

\begin{figure}
\centering
\includegraphics[width=.65\textwidth]{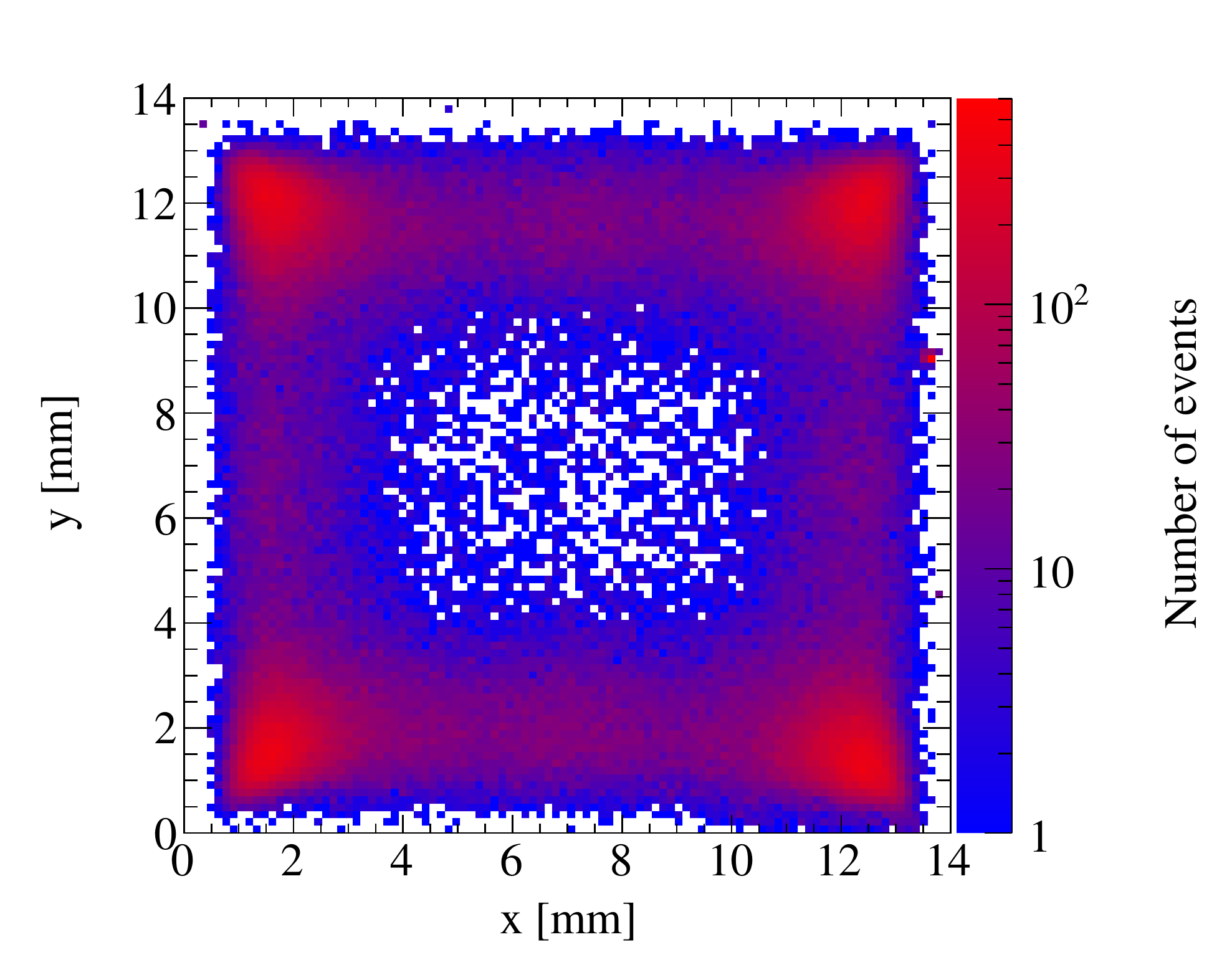}
\caption{Positions of X-ray like events up to \SI{10}{\keV} in the GridPix detector.}
\label{fig:background-positions}
\end{figure}

\begin{figure}
\centering
\includegraphics[width=.8\textwidth]{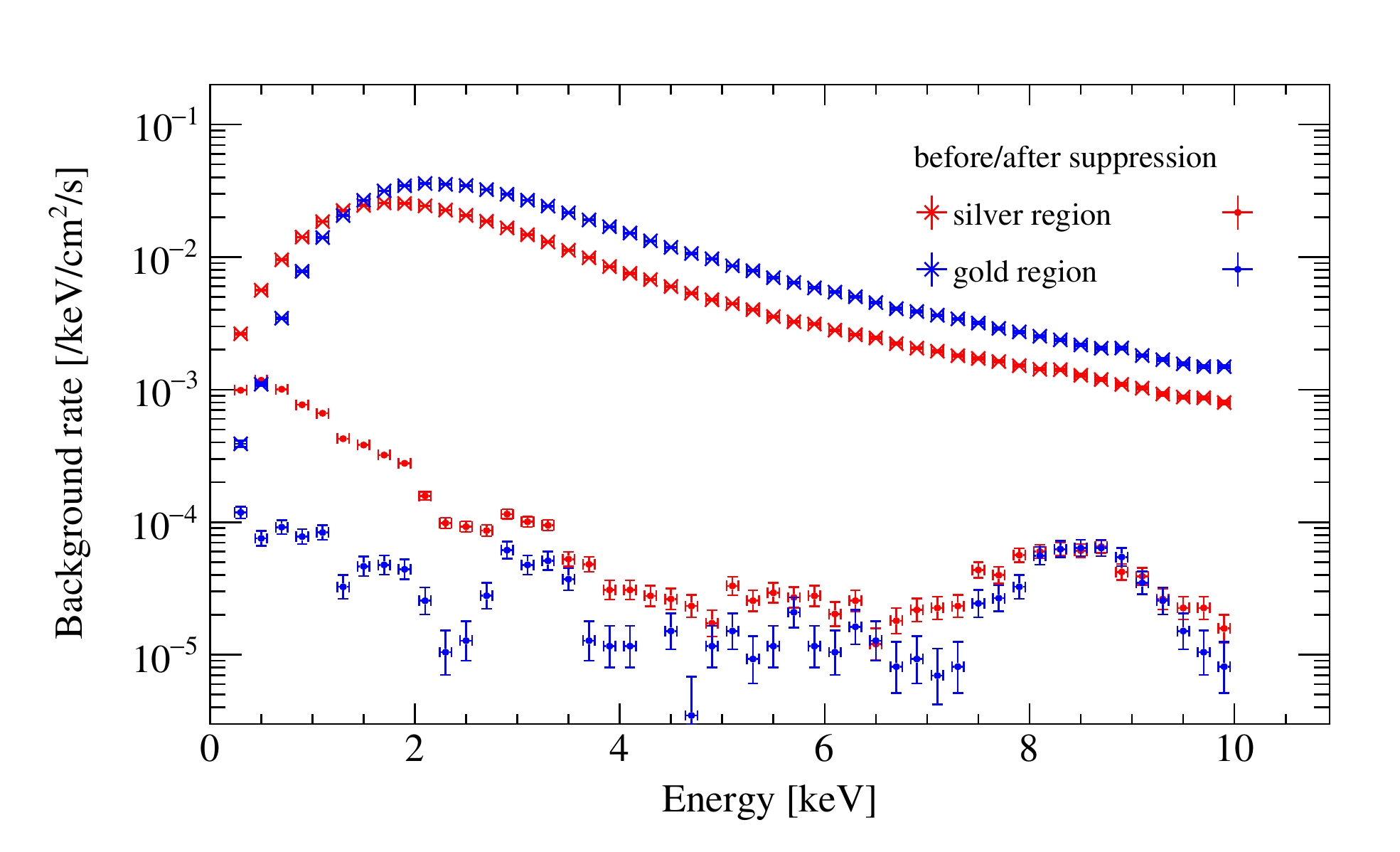}
\caption{Background rate before and after application of the likelihood based background suppression method for \textit{gold} and \textit{silver} region.}
\label{fig:background-rates}
\end{figure}

\Fref{fig:background-positions} shows the position of events passing the likelihood based background suppression for X-ray energies up to \SI{10}{\keV}. Only events within a radius of \SI{4.5}{\milli\metre} around the GridPix' center are fully contained on the chip with their eventshape and energy correctly reconstructed. The number of events, and therefore the background rate, is lowest in the central region. This is most likely caused by tracks only partially contained on the GridPix mimicking X-ray like, circular shaped events. Therefore two regions are defined: the \textit{gold} region comprising the innermost $\num{5}\times\SI{5}{\milli\metre\squared}$ of the chip and the \textit{silver} region which is a circle of \SI{4.5}{\milli\metre} radius around the chip center minus the \textit{gold} region. The resulting background rates before and after application of the likelihood based background suppression method are shown in \fref{fig:background-rates} for \textit{gold} and \textit{silver} region respectively. In the energy regime of \SIrange{0.2}{2}{\keV}, relevant for the detection of solar chameleons, a rate of less than \SI[per-mode=repeated-symbol]{e-4}{\per\keV\per\centi\metre\squared\per\second} is achieved in the \textit{gold} region which corresponds to an improvement of approximately one order of magnitude compared to the SDD used in CAST's previous chameleon search~\cite{anastassopoulos2015}.

\section{Analysis and results}

\begin{figure}
\centering
\includegraphics[width=.8\textwidth]{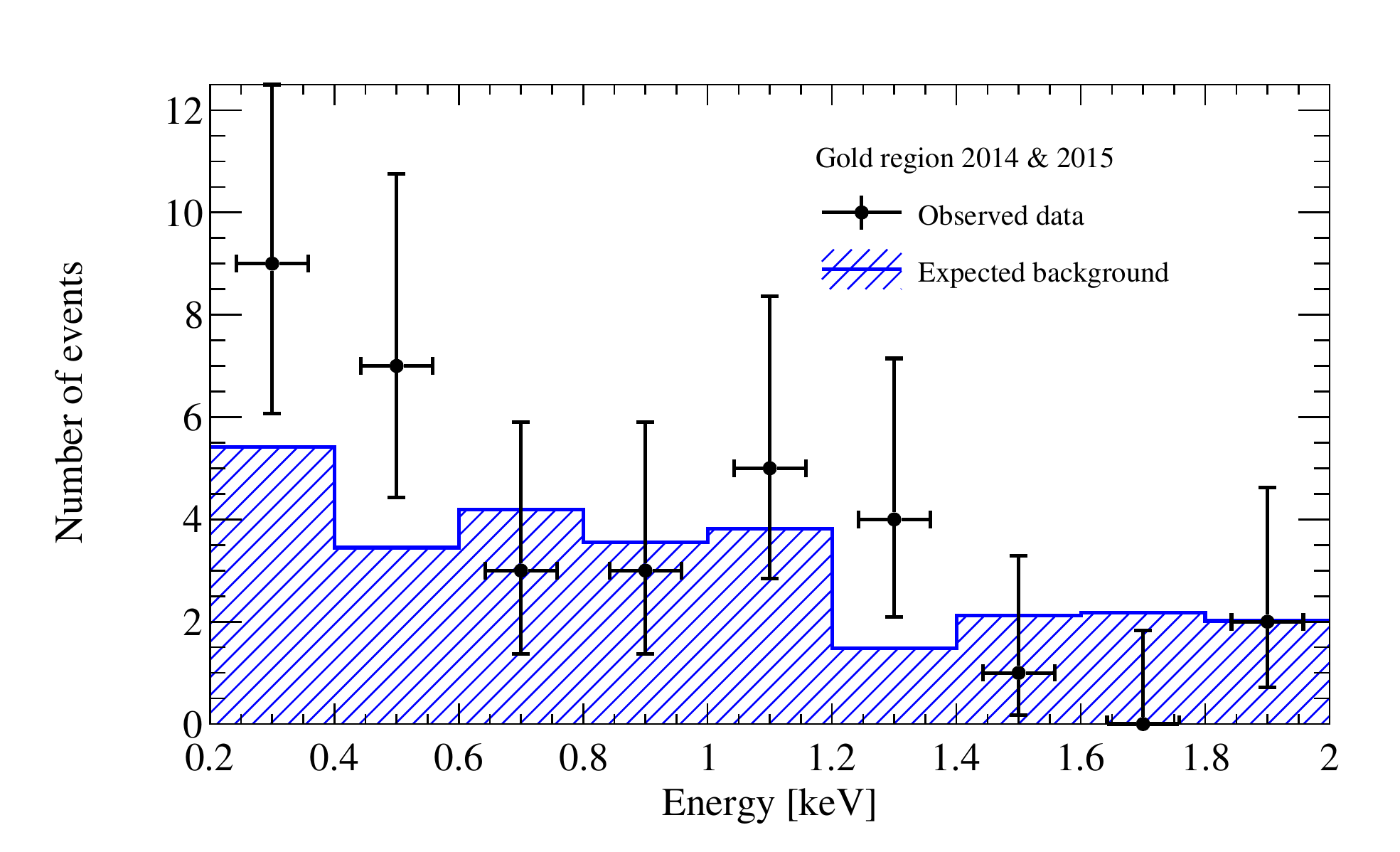}
\caption[Observed data points and background prediction in the gold region]{Observed data points and background prediction in the \textit{gold} region. Data points from the sunrise data set are almost perfectly compatible with the predicted background considering fluctuations within statistical uncertainties.}
\label{fig:sunrise-background-gold}
\end{figure}

\begin{figure}
\centering
\includegraphics[width=.8\textwidth]{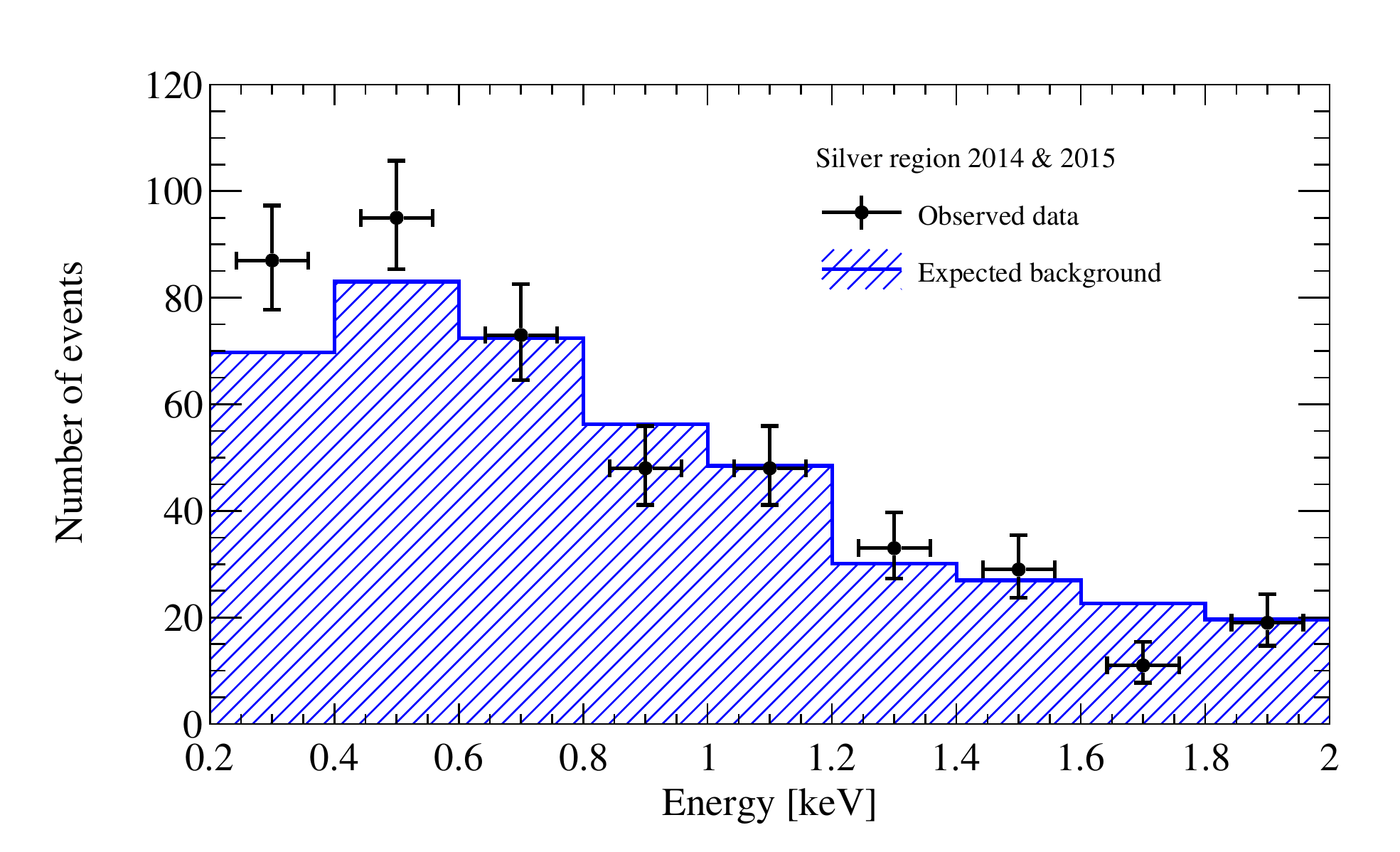}
\caption[Observed data points and background prediction in the silver region]{Observed data points and background prediction in the \textit{silver} region. Data points from the sunrise data set are well compatible with the predicted background considering fluctuations within statistical uncertainties.}
\label{fig:sunrise-background-silver}
\end{figure}

From the background rates in \fref{fig:background-rates} the expected number of background counts during the solar tracking can be extrapolated for the \textit{gold} and the \textit{silver} regions. In figures~\ref{fig:sunrise-background-gold} and~\ref{fig:sunrise-background-silver} the expected background counts are compared to the observed data points for an energy range from \SIrange{0.2}{2}{\keV}. Chameleons entering the evacuated cold bores of CAST first have to pass through the sunset detectors of CAST, in particular their lead shielding. Therefore, an energy threshold of \SI{0.2}{\keV} has been chosen which is well above the maximum chameleon effective mass in lead ($m_\text{eff}=\SI{135}{\eV}$ for $n=\num{1}$, $\beta_\text{m}=\num{e6}$)~\cite{anastassopoulos2015} so no absorption effects have to be taken into account.

The observed counts agree with the background prediction within statistical uncertainties, no excess over the prediction is observed. Hence, an upper bound on the chameleon photon coupling $\beta_\gamma$ can be calculated. Here, \texttt{TLimit} will be used for this purpose, the \texttt{ROOT} implementation of \texttt{mclimit}~\cite{junk1999}, which applies the likelihood ratio method to compute confidence levels $\text{CL}_b$ and $\text{CL}_{s+b}$ for the background-only and signal plus background hypotheses, respectively. The signal hypothesis is then tested using $\text{CL}_s = \text{CL}_{s+b}/\text{CL}_b$~\cite{read2002}. \texttt{TLimit} computes observed as well as expected confidence levels from given expected background, expected signal and observed data histograms; statistical as well as systematic uncertainties can be included.

\begin{figure}
\centering
\includegraphics[width=.8\textwidth]{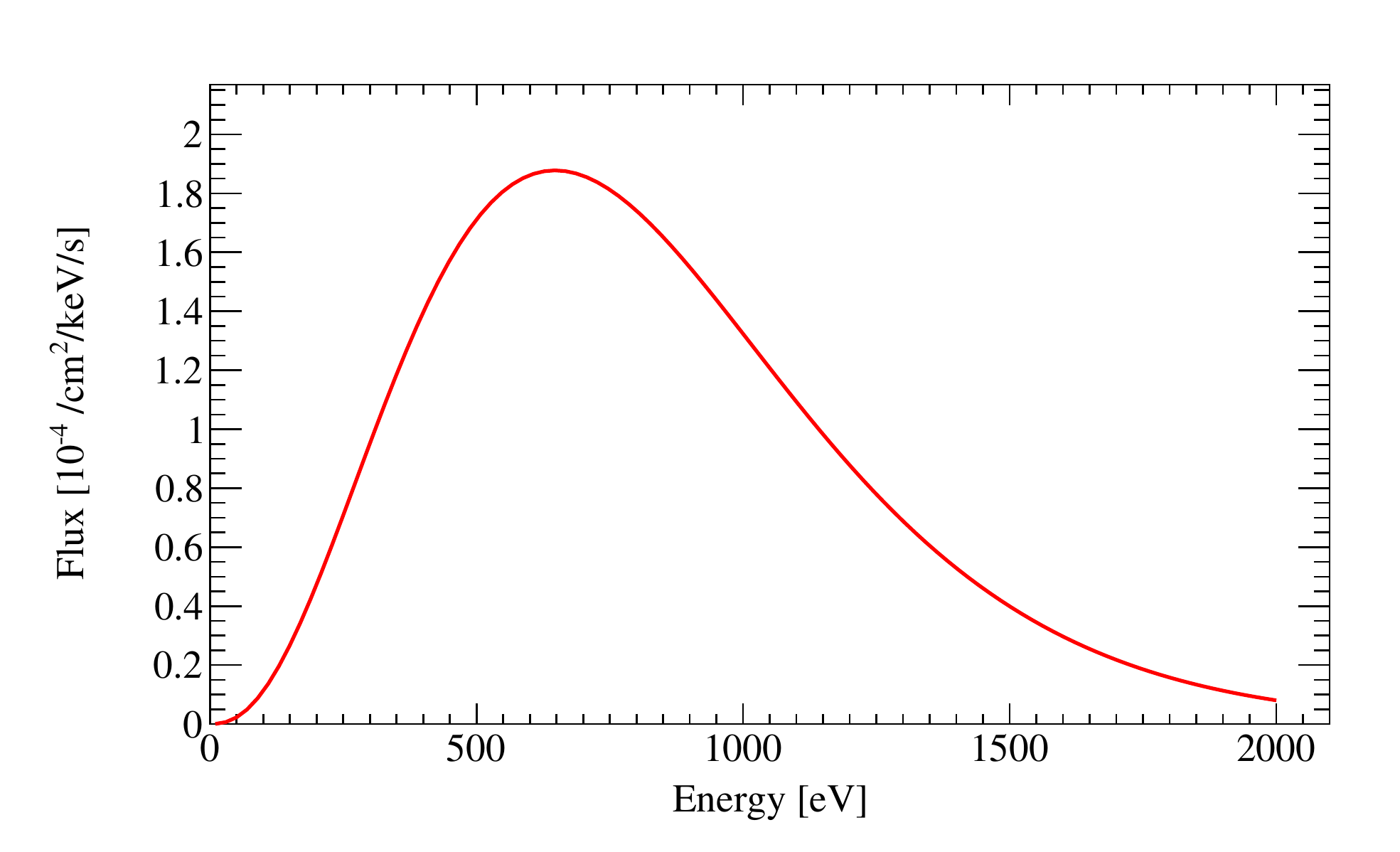}
\caption{Photon flux from solar chameleons reconverting into photons inside the CAST magnet for $\beta_\gamma=\beta_\gamma^\text{sun}$ in the case of non-resonant chameleon production. The aperture of the CAST magnet is not taken into account.}
\label{fig:chameleon-flux-cast}
\end{figure}

\begin{figure}
\centering
\includegraphics[width=.65\textwidth]{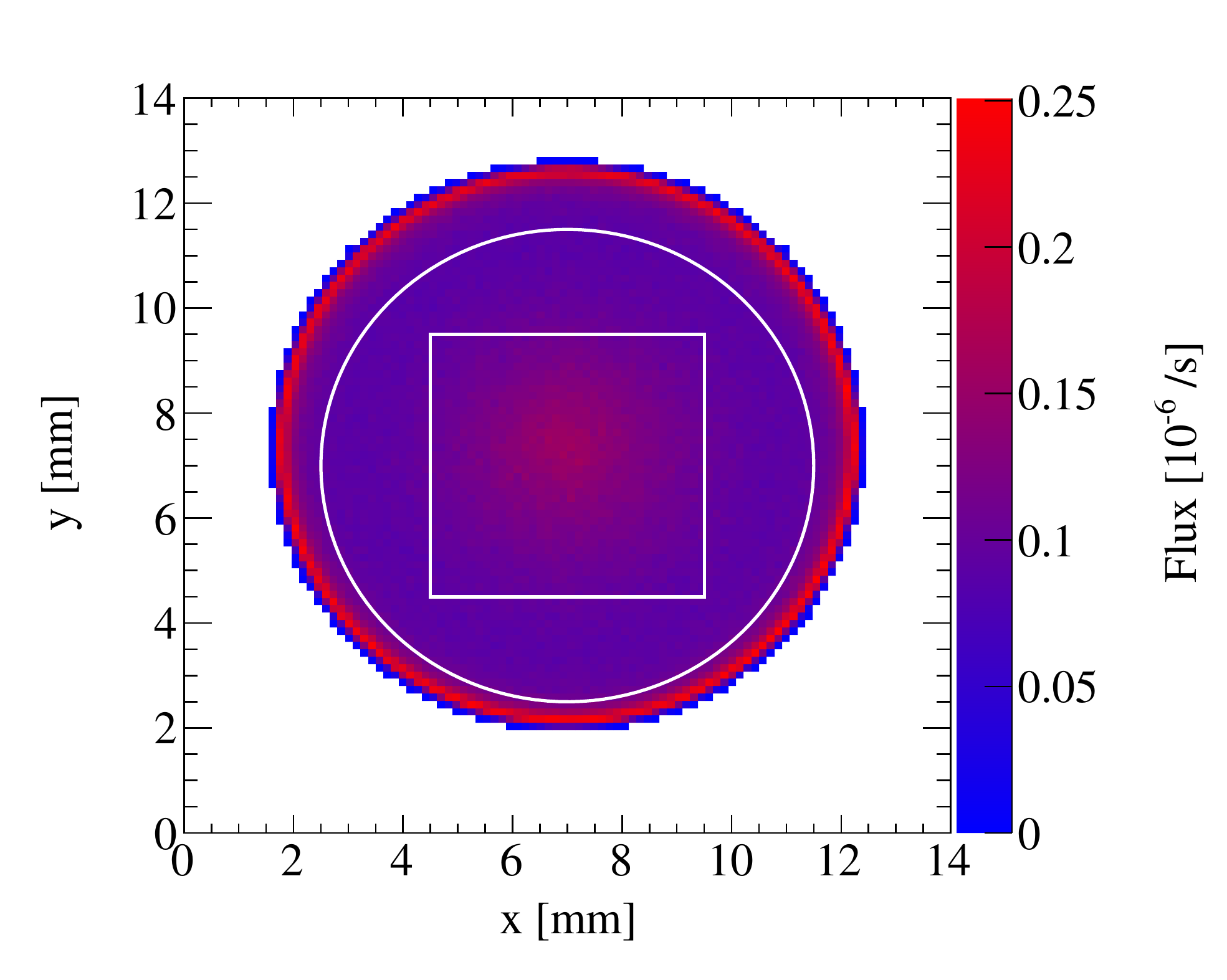}
\caption{Chameleon image of the Sun as observable at the GridPix detector's position taking into account the geometry of the CAST magnet as well as the imaging of the X-ray telescope. The square \textit{gold} and the circular \textit{silver} regions are indicated with white lines.}
\label{fig:chameleon-image}
\end{figure}

The expected signal in the two selected regions for a given chameleon photon coupling $\beta_\gamma$ is derived from the photon flux originating from solar chameleons converted inside the CAST magnet. The latter one is obtained by multiplying the solar chameleon flux with the conversion probability \eref{prop} using $B=\SI{9}{\tesla}$ and $l=\SI{9.26}{\metre}$, and shown in \fref{fig:chameleon-flux-cast} for $\beta_\gamma=\beta_\gamma^\text{sun}$. As the angular size of the solar tachocline (\SI{6.5}{\milli\radian} for a sphere with a diameter of \SI{0.7}{\solarradius}) is larger than the aperture of the CAST magnet not all possible chameleon trajectories which can reach the detector through the X-ray telescope see the full length of the magnetic field. This has to be taken into account in addition to the imaging of the X-ray telescope, especially its angular behavior and its efficiency (transmission) as function of X-ray energy. The X-ray telescope's efficiency drops (approximately) linearly with the off-axis angle of an incoming X-ray beam; for X-ray energies below \SI{2}{\keV} the efficiency drops to \SI{62.2}{\percent} of the on-axis value at an off-axis angle of \SI{10}{\arcmin}~\cite{kuster2007,friedrich1998}. The geometry of the CAST magnet bore and the imaging through the X-ray telescope are modeled in a small, simplified ray-tracing simulation similar to the one used in~\cite{anastassopoulos2015}. As a result the chameleon image of the Sun as observable at the detector's position can be computed as it is shown in \fref{fig:chameleon-image}. The ring-like shape originates from the production in the tachocline. The \textit{gold} and \textit{silver} regions are indicated in the image, about a third of the flux is lost outside the \text{silver} region. Despite the higher background level the \textit{silver} region still contributes to the observed upper bound on the chameleon photon coupling as it receives approximately the same flux as the \textit{gold} region.

\begin{figure}
\centering
\includegraphics[width=.8\textwidth]{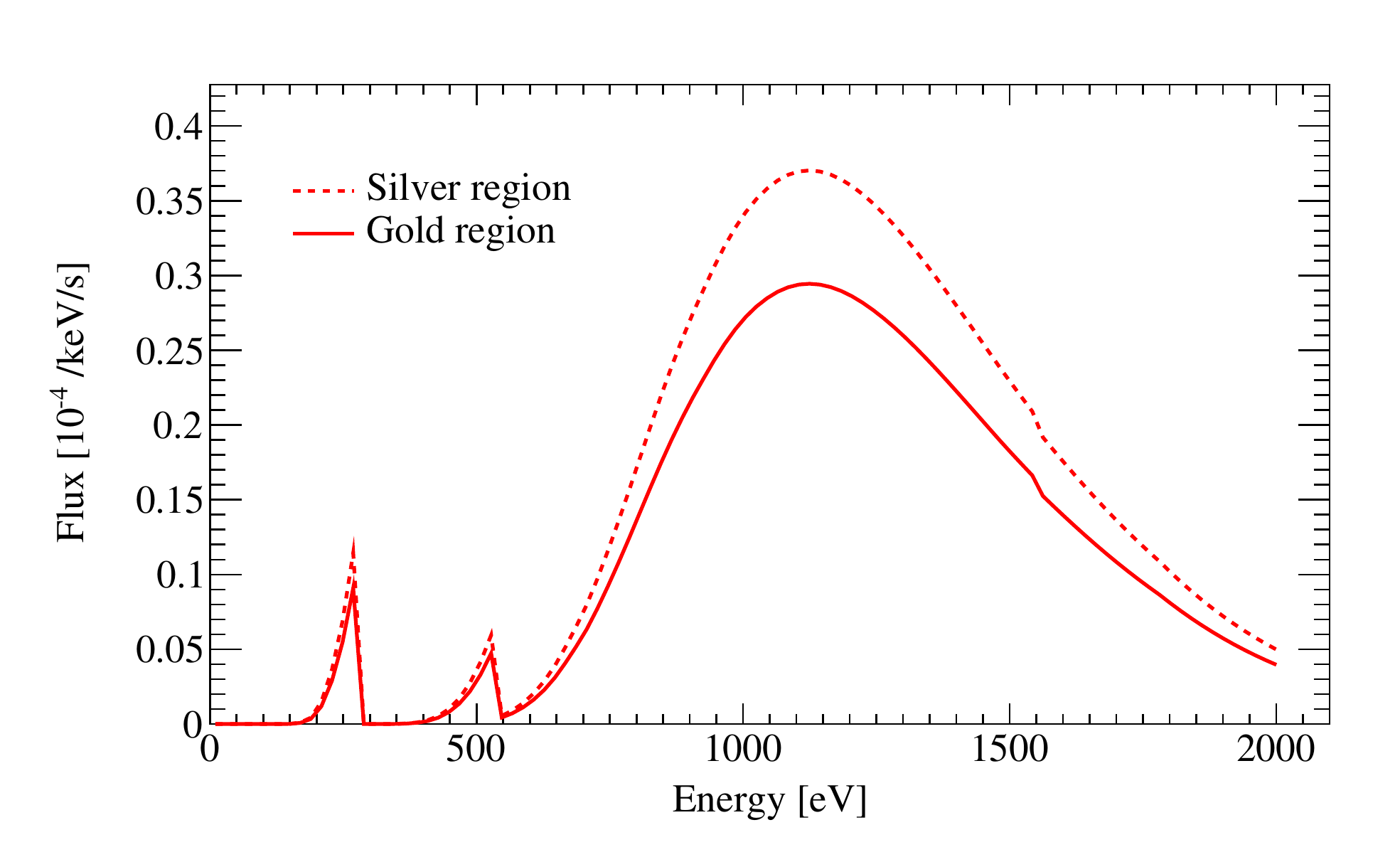}
\caption{Expected signal from solar chameleons observable with the GridPix detector in \textit{gold} and \textit{silver} region for a chameleon photon coupling $\beta_\gamma=\beta_\gamma^\text{sun}$.}
\label{fig:signal}
\end{figure}

To get the expected number of signal counts, in addition to the total solar tracking time, the detection efficiency of the GridPix detector (see \fref{fig_transparency}) and the software efficiency of the background suppression method are included. The resulting, observable solar chameleon spectra for both regions are shown in \fref{fig:signal} for a a chameleon photon coupling $\beta_\gamma=\beta_\gamma^\text{sun}$. The influence of the detector window on the shape of the chameleon spectrum is clearly visible. Especially, when comparing to the flux entering the XRT (see \fref{fig:chameleon-flux-cast}). The expected signal for different values of $\beta_\gamma$ can be obtained by rescaling the spectra according to their dependence on $\beta_\gamma^4$.

For the background prediction the statistical uncertainty on each bin is included in the analysis with \texttt{TLimit} while for the signal prediction different sources for systematic uncertainties are included; these sum up to a total systematic uncertainty on the expected signal of \SI{7.2}{\percent} (\SI{12.4}{\percent}) for the \textit{gold} (\textit{silver}) region. The individual contributions to the systematic uncertainty are listed in \tref{tab:sys} and were estimated by varying the corresponding parameters of the signal computation within reasonable ranges.

\begin{table}
\centering
\scalebox{1.0}{
\begin{tabular}{r|cc}
&\multicolumn{2}{c}{systematic uncertainty}\\
chip region&gold&silver\\
\hline
pointing accuracy&\SI{3.0}{\percent}&\SI{9.0}{\percent}\\
detector alignment&\SI{1.0}{\percent}&\SI{5.0}{\percent}\\
XRT off-axis behavior&\SI{1.5}{\percent}&\SI{3.0}{\percent}\\
XRT on-axis transmission&\multicolumn{2}{c}{\SI{1.8}{\percent}}\\
differential window transmission&\multicolumn{2}{c}{\SI{1.7}{\percent}}\\
detector window transmission&\multicolumn{2}{c}{\SI{3.9}{\percent}}\\
detector window optical transparency&\multicolumn{2}{c}{\SI{2.0}{\percent}}\\
detector gas absorption&\multicolumn{2}{c}{\SI{0.1}{\percent}}\\
software efficiency&\multicolumn{2}{c}{\SI{3.7}{\percent}}\\
\hline
total&\SI{7.2}{\percent}&\SI{12.4}{\percent}
\end{tabular}
}
\caption{List of estimated systematic uncertainties on the expected signal. The uncertainties resulting from imaging effects differ for \textit{gold} and \textit{silver} region and are therefore stated separately where necessary.}
\label{tab:sys}
\end{table}

\begin{figure}
\centering
\includegraphics[width=.8\textwidth]{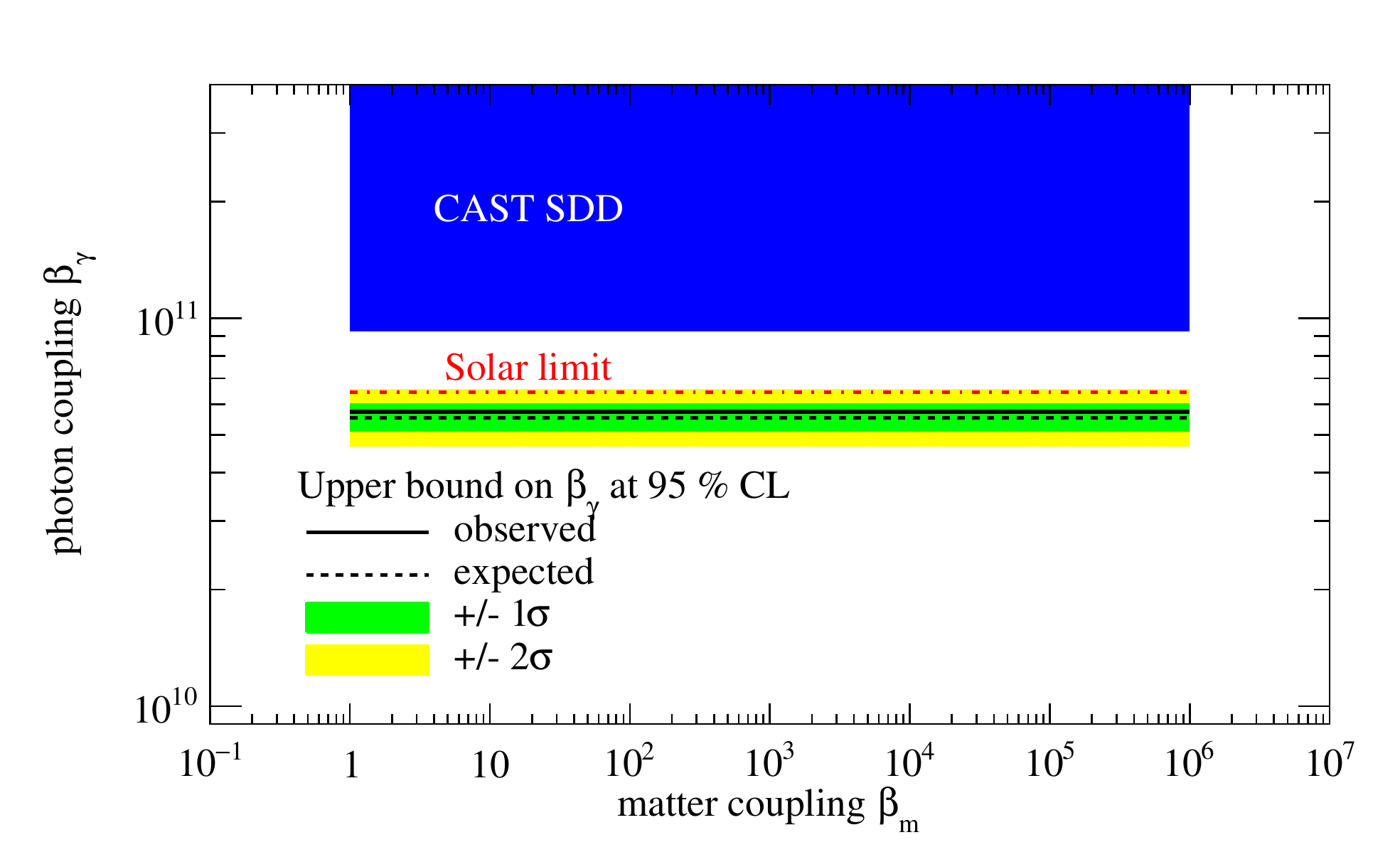}
\caption{Exclusion plot showing the observed upper bound on the chameleon photon coupling $\beta_\gamma$ obtained from the measurements with the GridPix detector in 2014 and 2015. The previous upper bound achieved by CAST using an SDD~\cite{anastassopoulos2015} is depicted in blue. The observed upper bound is shown as solid black line, the expected value as dashed black line with the $\pm1\sigma$ and $\pm2\sigma$ shown in green and yellow respectively. The upper limit given by the solar luminosity bound is shown as dot-dashed line in red. Only non-resonant chameleon production is taken into account.}
\label{fig:chameleon-limit-observed-cast}
\end{figure}

With these inputs, the expected \SI{95}{\percent} confidence level upper bound on the chameleon photon coupling is computed to be
\begin{equation}
\beta_\gamma<\left( 5.53^{+0.52}_{-0.43} \right) \times 10^{10}
\end{equation}
for $\num{1}<\beta_\text{m}<\num{e6}$ which is an improvement compared to our previous result by a factor of about two, as illustrated in \fref{fig:chameleon-limit-observed-cast}. Here, the uncertainty specifies the 1$\sigma$ range for experimental outcomes of hypothetical 
background-only experiments. The observed \SI{95}{\percent} confidence level upper bound on the chameleon photon coupling is
\begin{equation}
\beta_\gamma<\num{5.74e10}
\end{equation}
for non-resonant chameleon production ($\num{1}<\beta_\text{m}<\num{e6}$), which is indeed, as expected, below the upper limit given by the solar luminosity bound. 

\begin{figure}
\centering
\includegraphics[width=.8\textwidth]{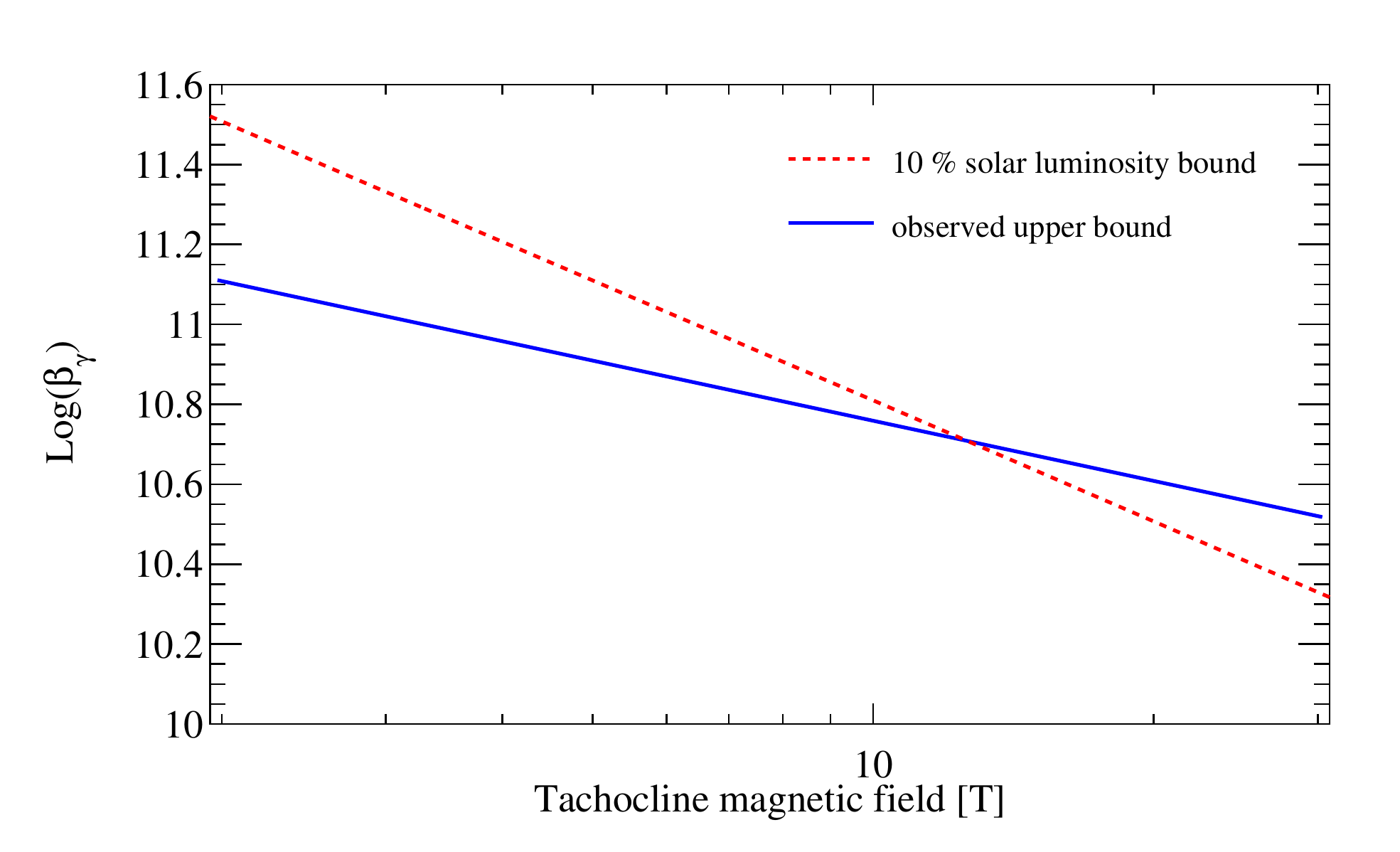}
\caption{Upper bound on $\beta_\gamma$ for different tachocline magnetic fields in blue. As comparison the \SI{10}{\percent} solar luminosity bound is drawn as dashed red line.}
\label{fig:limits-tachocline-field}
\end{figure}

Of course, the obtained result depends on the solar model considered. Here, we focused on the scenario with a magnetic field of \SI{10}{\tesla} in the tachocline. The uncertainty on the tachocline field is believed to be in the range of \SIrange{4}{25}{\tesla}~\cite{weber2013,caligari1995,antia2003}. Hence the CAST limit on the chameleon photon coupling can be shifted by a factor of about $2.5^{1/2}$ up or down as illustrated in \fref{fig:limits-tachocline-field}. The limit obtained with the GridPix detector is below the solar luminosity bound for values of tachocline magnetic fields up to \SI{12.5}{\tesla}.

As for our previous result, we have studied the influence of the tachocline position and width on the observed upper bound on the chameleon photon coupling. The tachocline has been shifted down to \SI{0.66}{\solarradius} and its width changed from \SI{0.01}{\solarradius} to \SI{0.04}{\solarradius}. Additionally, a linearly decreasing magnetic field (\SI{10}{\tesla} at \SI{0.7}{\solarradius} down to \SI{0}{\tesla} at \SI{0.8}{\solarradius}) has been considered. The changes to the bound on $\beta_\gamma$ can be found in \tref{tab:limits-tachocline}. For most of the scenarios the observed bound is below the solar luminosity bound and we found, that in general, irrespectively of the astrophysics of the tachocline, $\beta_\gamma<\num{6e10}$ is satisfied.

\begin{table}
\centering
\scalebox{1.0}{
\begin{tabular}{ll|cc|c}
\multicolumn{2}{c|}{Tachocline}&\multicolumn{2}{c|}{$\beta_\gamma$ at \SI{95}{\percent} CL}&\\
position [$R_\odot$]& width [$R_\odot$]&expected&observed&$\beta_\gamma^\text{sun}$\\
\hline
\rule{0pt}{15pt}\num{0.7}&\num{0.01}&$\left( 5.53^{+0.52}_{-0.43} \right) \times 10^{10}$&\num{5.74e10}&\num{6.46e10}\\
\rule{0pt}{15pt}\num{0.66}&\num{0.01}&$\left( 4.94^{+0.45}_{-0.39} \right) \times 10^{10}$&\num{4.98e10}&\num{5.89e10}\\
\rule{0pt}{15pt}\num{0.66}&\num{0.04}&$\left( 3.54^{+0.31}_{-0.28} \right) \times 10^{10}$&\num{3.58e10}&\num{2.95e10}\\
\rule{0pt}{15pt}\num{0.7}&\num{0.1} linear&$\left( 4.19^{+0.38}_{-0.33} \right) \times 10^{10}$&\num{4.36e10}&\num{4.47e10}
\end{tabular}
}
\caption{Upper bound on $\beta_\gamma$ derived from the measurements with the GridPix detector for different solar models, all using the \SI{10}{\percent} solar luminosity bound.}
\label{tab:limits-tachocline}
\end{table}

\section{Discussion}

\begin{figure}
\centering
\includegraphics[width=.8\textwidth]{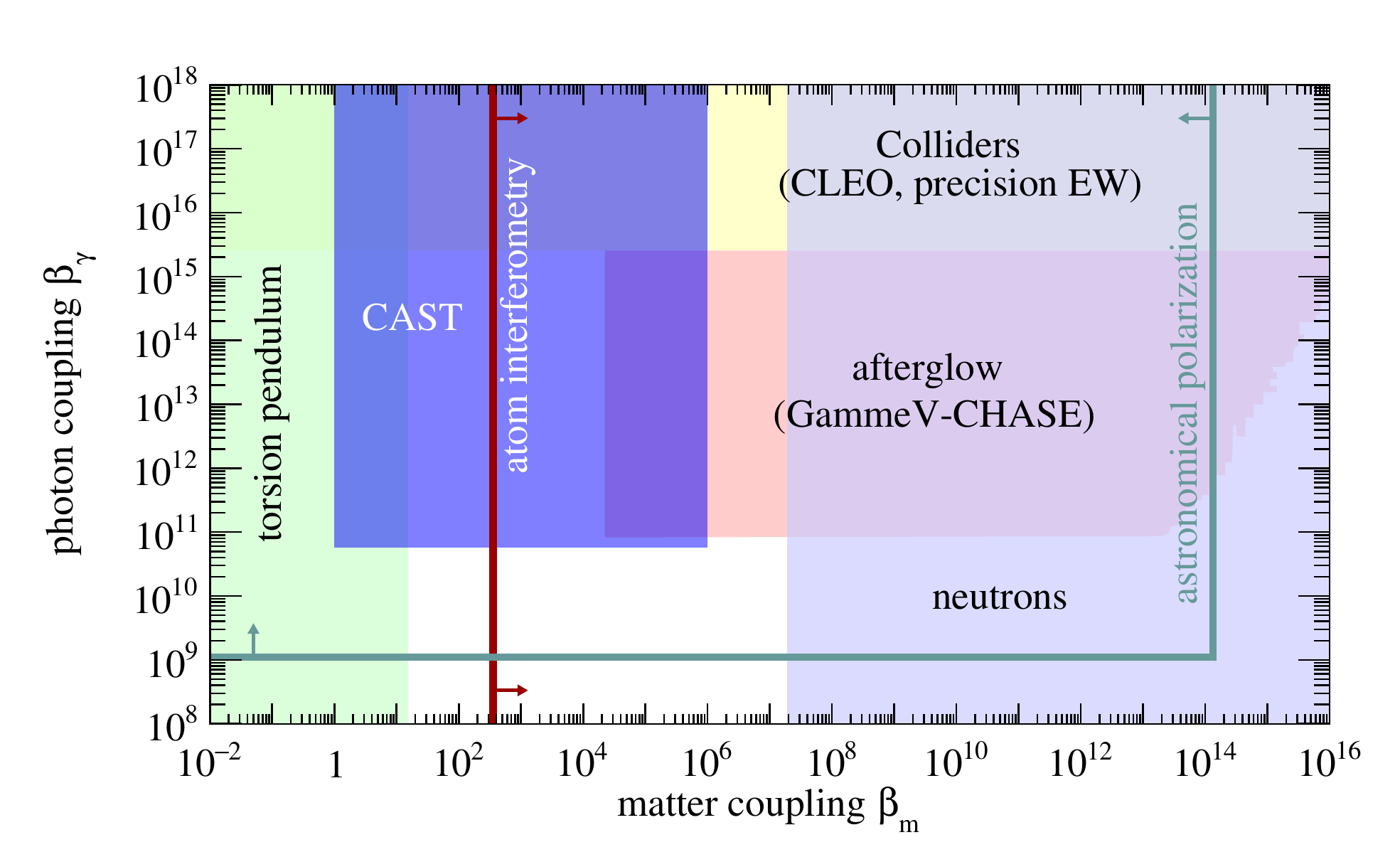}
\caption[Exclusion region for chameleons in the $\beta_\gamma$-$\beta_m$ plane]{Exclusion region for chameleons in the $\beta_\gamma$-$\beta_m$ plane, achieved by CAST in 2014 and 2015 using the data taken with the GridPix detector (blue). Also shown are the bounds set by torsion pendulum tests (green)~\cite{upadhye2012}, neutron interferometry measurements (lilac)~\cite{lemmel2015}, CHASE (pale orange)~\cite{steffen2010} and collider experiments (yellow)~\cite{brax2009}. The bounds of the atom-interferometry technique~\cite{hamilton2015,jaffe2017} and the astronomical polarization~\cite{burrage2009} are represented with lines.}
\label{fig:chameleon-exclusion-plot}
\end{figure}

The chameleon parameter space is spanned by three parameters: the chameleon matter coupling constant $\beta_\text{m}$, the (effective) chameleon photon coupling constant $\beta_\gamma$ and a discrete index $n$ which defines the Dark Energy model considered. Our improved result for the upper limit on the chameleon photon coupling $\beta_\gamma$ is presented in \fref{fig:chameleon-exclusion-plot} for $n=1$ along with other experimental bounds. Some of these are only sensitive to the chameleon matter coupling and, therefore result in vertical lines in the shown exclusion plot. While torsion pendulum tests of presence of new scalar forces lead to a lower bound on the chameleon matter coupling $\beta_\text{m}$ (in green)~\cite{upadhye2012}, neutron interferometry (lilac)~\cite{lemmel2015} and the atom-interferometry technique (dark red line)~\cite{hamilton2015,jaffe2017} lead to an upper bound where the latter provides the strongest bound. A large upper bound on the chameleon photon coupling is provided by precision tests of the Standard Model~\cite{brax2009}, these kinds of tests are sensitive only to the coupling to gauge bosons, in this case photons. From astrophysics an upper bound can be derived through analysis of the polarisation of the light coming from astronomical objects~\cite{burrage2009}.

While our result only considers non-resonant chameleon production in the Sun and therefore only provides an upper bound for the chameleon photon coupling constant for a chameleon matter coupling up to $\beta_\text{m}<\num{e6}$, the CHASE experiment~\cite{steffen2010} is sensitive to $\beta_\gamma$ up to $\beta_\text{m}\sim\num{e14}$. Taking into account the strong upper and lower bounds provided by the atom-interferometry technique and torsion pendulum experiments only the region of $\num{15.33}<\beta_\text{m}<\num{3.57e2}$ has not yet been excluded on the $\beta_\text{m}$ axis. This small region is further and significantly reduced by our improved upper bound on the chameleon photon coupling $\beta_\gamma<\num{5.74e10}$ leaving only a small part of the parameter space open for the $n=1$ scenario.

As already shown in CAST's previous solar chameleon search~\cite{anastassopoulos2015}, also our improved result is to a large extent insensitive to $n$, as visible from the results for different values of $n$ listed in \tref{tab:limits-indexn}. This insensitivity to $n$ is caused by the restriction to non-resonant chameleon production in the Sun.

\begin{table}
\centering
\scalebox{1.0}{
\begin{tabular}{c|cc}
&\multicolumn{2}{c}{$\beta_\gamma$ at \SI{95}{\percent} CL}\\
Index $n$&expected&observed\\
\hline
\rule{0pt}{15pt}\num{1}&$\left( 5.53^{+0.52}_{-0.43} \right) \times 10^{10}$&\num{5.74e10}\\
\rule{0pt}{15pt}\num{2}&$\left( 5.51^{+0.51}_{-0.43} \right) \times 10^{10}$&\num{5.69e10}\\
\rule{0pt}{15pt}\num{4}&$\left( 5.49^{+0.49}_{-0.44} \right) \times 10^{10}$&\num{5.67e10}\\
\rule{0pt}{15pt}\num{6}&$\left( 5.50^{+0.50}_{-0.44} \right) \times 10^{10}$&\num{5.67e10}
\end{tabular}
}
\caption[Upper bound on $\beta_\gamma$ for different values of the index $n$]{Upper bound on $\beta_\gamma$ derived from the measurements with the GridPix detector at CAST for different values of the index $n$ which defines the chameleon model.}
\label{tab:limits-indexn}
\end{table}

In addition, the influence of uncertainties in the solar model assumptions have been studied. For example, if the solar luminosity bound is reduced by one order of magnitude, $\beta_\gamma^\text{sun}$ is reduced by a factor $\sqrt{10}$ while the observed limit remains unchanged, thus weakening the limit with respect to the solar luminosity bound. Uncertainty of the tachocline size and position as well as the radial field strength and distribution, may lead to a change of the observed limit by a factor \num{1.6} (see \tref{tab:limits-tachocline}) following a conservative approach. Similarly, also the uncertainty on the tachocline magnetic field strength give an uncertainty on the observed upper bound for $\beta_\gamma$ corresponding to a factor \num{1.6} up or down considering magnetic fields in the range of \SIrange{4}{25}{\tesla}, see also \fref{fig:limits-tachocline-field}.

\section{Conclusions}

Summarizing, the first upper bound derived by the experimental approach using a magnetic helioscope, on the chameleon photon coupling $\beta_\gamma$ derived by CAST using an SDD without X-ray optics has been significantly improved by utilizing a novel GridPix detector in combination with the MPE XRT, now giving an upper bound of $\beta_\gamma<\num{5.74e10}$ for $\num{1}<\beta_\text{m}<\num{e6}$, which for the first time in CAST allows surpassing the solar luminosity bound. Together with other experimental bounds, this restricts the chameleon parameter space to the region confined by ${\num{15.33}<\beta_\text{m}<\num{3.57e2}}$ and $\beta_\gamma<\num{5.74e10}$ for $n=1$.

This result may be further improved in the near future by a new and upgraded GridPix detector currently operated at CAST now covering a larger active area, implementing the analogue Grid signal as well as two veto scintillators which along with improved background suppression algorithms should lead to a significant reduction in background rate. Along with the first time usage of ultrathin silicon nitride windows that increase the detector's X-ray detection efficiency in the sub-\SI{}{\keV} range, this should lead to a further improved sensitivity.

\section{Acknowledgments}

We thank CERN for hosting the experiment and for technical support to operate the magnet and cryogenics. 

We acknowledge support from NSERC (Canada), MSE (Croatia) and Croatian Science Foundation under the project IP-2014-09-3720, CEA (France), BMBF (Germany) under the grant numbers 05 CC2EEA/9 and 05 CC1RD1/0 and DFG (Germany) under grant numbers HO 1400/7-1 and EXC-153, GSRT (Greece), NSRF: Heracleitus II, RFFR (Russia), the Spanish Ministry of Economy and Competitiveness (MINECO) under Grants No.\ FPA2011-24058 and No.\ FPA2013-41085-P (grants partially funded by the European Regional Development Fund, ERDF/FEDER), the European Research Council
(ERC) under grant ERC-2009-StG-240054 (T-REX), Turkish Atomic Energy Authority (TAEK), NSF (USA) under Award No.\ 0239812, NASA under the grant number NAG5-10842, the University of Rijeka under grant number 13.12.2.2.09, and IBS (Korea) with code IBS-R017-D1-2017-a00. Part of this
work was performed under the auspices of the U.S.\ Department of Energy by Lawrence Livermore National Laboratory under Contract No.\ DE-AC52-07NA27344.

\section*{References}

\bibliography{main}

\providecommand{\newblock}{}
\begin{thebibliography}{10}
\expandafter\ifx\csname url\endcsname\relax
  \def\url#1{{\tt #1}}\fi
\expandafter\ifx\csname urlprefix\endcsname\relax\def\urlprefix{URL }\fi
\providecommand{\eprint}[2][]{\url{#2}}

\bibitem{khoury2004}
Khoury J and Weltman A 2004 {\em Phys. Rev. Lett.\/} {\bf 93} 171104

\bibitem{khoury2004a}
Khoury J and Weltman A 2004 {\em Phys. Rev.\/} D {\bf 69} 044026

\bibitem{brax2004}
Brax P, van~de Bruck C, Davis A~C, Khoury J and Weltman A 2004 {\em Phys.
  Rev.\/} D {\bf 70} 123518

\bibitem{joyce2015}
Joyce A, Jain B, Khoury J and Trodden M 2015 {\em Phys. Rep.\/} {\bf 568} 1--98

\bibitem{burrage2018}
Burrage C and Sakstein J 2018 {\em Living Rev. Relativ.\/} {\bf 21} 1

\bibitem{brax2010}
Brax P and Zioutas K 2010 {\em Phys. Rev.\/} D {\bf 82} 043007

\bibitem{brax2012}
Brax P, Lindner A and Zioutas K 2012 {\em Phys. Rev.\/} D {\bf 85} 043014

\bibitem{zioutas1999}
Zioutas K {\em et~al.\/} 1999 {\em Nucl. Instr. Meth. Phys. Res.\/} A {\bf 425}
  480--487

\bibitem{anastassopoulos2015}
Anastassopoulos V {\em et~al.\/} (CAST collaboration) 2015 {\em Phys. Lett.\/}
  B {\bf 749} 172--180

\bibitem{kuster2007}
Kuster M {\em et~al.\/} 2007 {\em New J. Phys.\/} {\bf 9} 169

\bibitem{krieger2013}
Krieger C, Kaminski J and Desch K 2013 {\em Nucl. Instr. Meth. Phys. Res. A\/}
  {\bf 729} 905--909

\bibitem{krieger2017}
Krieger C, Kaminski J, Lupberger M and Desch K 2017 {\em Nucl. Instr. Meth.
  Phys. Res.\/} A {\bf 867} 101--107

\bibitem{baum2014}
Baum S, Cantatore G, Hoffmann D, Karuza M, Semertzidis Y, Upadhye A and Zioutas
  K 2014 {\em Phys. Lett.\/} B {\bf 739} 167--173

\bibitem{anastassopoulos2017}
Anastassopoulos V {\em et~al.\/} (CAST collaboration) 2017 {\em Nat. Phys.\/}
  {\bf 13} 584--590

\bibitem{zioutas2005}
Zioutas K {\em et~al.\/} (CAST collaboration) 2005 {\em Phys. Rev. Lett.\/}
  {\bf 94} 121301

\bibitem{andriamonje2007}
Andriamonje S {\em et~al.\/} (CAST collaboration) 2007 {\em J. Cosmol.
  Astropart. Phys.\/} {\bf 2007} 010

\bibitem{arik2009}
Arik E {\em et~al.\/} (CAST collaboration) 2009 {\em J. Cosmol. Astropart.
  Phys.\/} {\bf 2009} 008

\bibitem{arik2011}
Arik M {\em et~al.\/} (CAST collaboration) 2011 {\em Phys. Rev. Lett.\/} {\bf
  107} 261302

\bibitem{arik2014}
Arik M {\em et~al.\/} (CAST collaboration) 2014 {\em Phys. Rev. Lett.\/} {\bf
  112} 091302

\bibitem{llopart2007}
Llopart X, Ballabriga R, Campbell M, Tlustos L and Wong W 2007 {\em Nucl.
  Instr. Meth. Phys. Res.\/} A {\bf 581} 485--494

\bibitem{gullikson2010}
Gullikson E 2010 {X-Ray Interactions With Matter}
  \urlprefix\url{http://henke.lbl.gov/optical_constants/}

\bibitem{henke1993}
Henke B, Gullikson E and Davis J 1993 {\em At. Data Nucl. Data Tables\/} {\bf
  54} 181--342 ISSN 0092-640X

\bibitem{krieger2018}
Krieger C, Kaminski J, Vafeiadis T and Desch K 2018 {\em Nucl. Instr. Meth.
  Phys. Res.\/} A {\bf 893} 26--34

\bibitem{abernathy2008}
Abernathy J {\em et~al.\/} 2008 {MarlinTPC: A common software framework for TPC
  development} {\em 2008 IEEE Nuclear Science Symposium Conference Record\/} pp
  1704--1708

\bibitem{junk1999}
Junk T 1999 {\em Nucl. Instr. Meth. Phys. Res.\/} A {\bf 434} 435--443

\bibitem{read2002}
Read A 2002 {\em J. Phys. G: Nucl. Part. Phys\/} {\bf 28} 2693

\bibitem{friedrich1998}
Friedrich P, Br\"auninger H, Burkert W, D\"ohring T, Egger R, Hasinger G,
  Oppitz A, Predehl P and Tr\"umper J 1998 {X-ray tests and calibrations of the
  ABRIXAS mirror systems} {\em X-Ray Optics, Instruments, and Missions\/} vol
  3444 p 369

\bibitem{weber2013}
Weber M, Fan Y and Miesch M 2013 {\em Sol. Phys.\/} {\bf 287} 239--263

\bibitem{caligari1995}
Caligari P, Moreno-Insertis F and Sch\"ussler M 1995 {\em Astrophys. J.\/} {\bf
  441} 886--902

\bibitem{antia2003}
Antia H, Chitre S and Thompson M 2003 {\em Astron. Astrophys.\/} {\bf 399}
  329--336

\bibitem{upadhye2012}
Upadhye A 2012 {\em Phys. Rev.\/} D {\bf 86} 102003

\bibitem{lemmel2015}
Lemmel H, Brax P, Ivanov A, Jenke T, Pignol G, Pitschmann M, Potocar T,
  Wellenzohn M, Zawisky M and Abele H 2015 {\em Phys. Lett.\/} B {\bf 743}
  310--314

\bibitem{hamilton2015}
Hamilton P, Jaffe M, Haslinger P, Simmons Q, M\"uller H and Khoury J 2015 {\em
  Science\/} {\bf 349}(6250) 849--851

\bibitem{jaffe2017}
Jaffe M, Haslinger P, Xu V, Hamilton P, Upadhye A, Elder B, Khoury J and
  M\"uller H 2017 {\em Nat. Phys.\/} {\bf 13} 938--942

\bibitem{brax2009}
Brax P, Burrage C, Davis A, Seery D and Weltman A 2009 {\em J. High Energy
  Phys.\/} {\bf 09} 128

\bibitem{burrage2009}
Burrage C, Davis A and Shaw D 2009 {\em Phys. Rev.\/} D {\bf 79} 044028

\bibitem{steffen2010}
Steffen J, Upadhye A, Baumbaugh A, Chou A, Mazur P, Tomlin R, Weltman A and
  Wester W 2010 {\em Phys. Rev. Lett.\/} {\bf 105} 261803

\end{thebibliography}

\end{document}